\documentclass[a4paper,english,aps,floats,twocolumn,showpacs,nofootinbib]{revtex4}
\usepackage{pslatex}
\usepackage[T1]{fontenc}
\usepackage{graphicx}
\usepackage{epsfig}

\usepackage{calc}
\usepackage{ifthen}
\usepackage{subfigure}

\newcommand{\bea}{\begin{eqnarray}}
\newcommand{\eea}{\end{eqnarray}}

\newcommand{\nc}{\newcommand}
\nc{\renc}{\renewcommand}
\nc{\eqs}[2]{\mbox{Eqs.~(\ref{#1},\,\ref{#2})}}
\nc{\eq}[1]{\mbox{Eq.~(\ref{#1})}}
\nc{\figs}[2]{\mbox{Figs.~(\ref{#1},\,\ref{#2})}}
\nc{\fig}[1]{\mbox{Fig~.(\ref{#1})}}
\nc{\be}[1]{\begin{equation} \mbox{$\label{#1}$}}
\nc{\ee}{\vspace{0.1cm}\end{equation}}

\newcommand{\bean}{\begin{eqnarray*}}
\newcommand{\eean}{\end{eqnarray*}}

\def\GeV{{\rm \ GeV}}
\def\MeV{{\rm \ MeV}}

\def\TeV{{\rm \ TeV}}

\def\lae{\;^{<}_{\sim} \;} \def\gae{\; ^{>}_{\sim} \;}

\def\pbh{\hat{\phi}_{B}}

\def\tH{\hat{H}}

\nc{\npp}[3]{{\it  Nucl.\ Phys.\ }{{\bf #1} {(#2)} {#3}}}
\nc{\prdd}[3]{{\it  Phys.\ Rev.\ D\ }{{\bf #1} {(#2)} {#3}}}
\nc{\prll}[3]{{\it Phys.\ Rev.\ Lett.\ }{{\bf #1} {(#2)} {#3}}}
\nc{\pll}[3]{{\it  Phys.\ Lett.\ }{{\bf #1} {(#2)} {#3}}}

\begin{document}
\title{ Baryomorphosis: Relating the Baryon Asymmetry to the "WIMP Miracle"
}
\author{John McDonald}
\email{j.mcdonald@lancaster.ac.uk}
\affiliation{Cosmology and Astroparticle Physics Group, University of
Lancaster, Lancaster LA1 4YB, UK}
\begin{abstract}

     We present a generic framework, {\it baryomorphosis}, which modifies the baryon asymmetry to be naturally of the order of a typical thermal relic WIMP density. We consider a simple scalar-based model to show how this is possible. 
This model introduces a sector in which a large initial baryon asymmetry is injected into particles ("annihilons"), $\phi_{B}$, $\pbh$, of mass $\sim 100 \GeV - 1 \TeV$. $\phi_{B} \; \pbh$  annihilations convert the initial 
$\phi_{B}$, $\pbh$ asymmetry to a final asymmetry with a thermal relic WIMP-like density. This subsequently decays to a conventional baryon asymmetry whose magnitude is naturally related to the density of thermal relic WIMP dark matter. In this way the two coincidences of baryons and dark matter i.e. why their densities are similar to each other and why they are both similar to a WIMP thermal relic density (the "WIMP miracle"), may be understood. The model may be tested by the production of annihilons at colliders.

 \end{abstract}
\maketitle

\section{Introduction}

       The origin of the baryon asymmetry is one of the fundamental questions of cosmology. Several mechanisms have been proposed: out-of-equilibrium decay of GUT bosons, electroweak baryogenesis, Affleck-Dine baryogenesis, varieties of leptogenesis and many others. However, these mechanisms do not address the similarity of the mass density of baryons and dark matter, $\rho_{B}/\rho_{DM} \approx 1/6$. This has motivated several models which attempt to unify the 
origin of baryons and of dark matter, which typically relate the baryon asymmetry to a dark matter asymmetry 
\cite{bdm0,bdm1}. Supersymmetric models which relate baryon number and dark matter via conserved R-parity have also been proposed \cite{susybdm0,susybdm1}.  We refer to these as 'direct' models, as they directly relate
the {\it number} density of the baryons to that of dark matter. As a result, the mass of the dark matter particle has to be light, typically $\sim 1-10$ GeV. (Exceptions have been suggested in \cite{x1,xo,x2}.) Recently, a number of models based on $\GeV$-mass asymmetric dark matter have been proposed \cite{bdmr}.

    It is also possible to relate the baryon and dark matter densities {\it indirectly}, by having similar mechanisms for the generation of the observed densities. An MSSM-based example was in given \cite{jm1}, where the baryon number originates from Affleck-Dine leptogenesis while the dark matter density is due to coherently-oscillating right-handed sneutrinos. Both the baryon and dark matter density are from flat directions with the same dimension, in which case their mass densities are related. The advantage of such indirect models is that they allow a wider range of dark matter particle mass, which is not constrained by a tight relation between the baryon and dark matter number densities.

  However, most models which address the baryon to dark matter ratio problem sacrifice another coincidence, the similarity of the observed dark matter density to a typical thermal relic WIMP density, the so-called "WIMP miracle".  Ideally we would construct a framework which relates the baryon and dark matter densities while retaining the connection to the thermal relic WIMP density. This could be achieved if the dark matter is a conventional WIMP while the {\it final} baryon asymmetry is determined by a process which is similar to the process responsible for thermal WIMP densities i.e. the freeze-out of annihilations of dark matter particles which have an annihilation cross-section of broadly weak-interaction strength. 
The goal of this paper is to present a model which demonstrates such a process. We emphasize that the framework is more general than the implementation we present here. We do not address the origin of the initial baryon asymmetry, but instead concentrate on how this can be modified to become a thermal WIMP-like density, hence "baryomorphosis" rather than "baryogenesis". We will show that a generic feature of such models are new particles, which we term "annihilons", which allow annihilation of the initial baryon asymmetry down to a thermal WIMP-like density. Depending on their Standard Model (SM) gauge charges, these could provide a distinctive signature of the model at colliders.

\section{Baryomorphosis: An Illustrative Example}   

   To show how it is possible to generate a baryon asymmetry which is related to a thermal relic WIMP density, we will construct two versions of a simple scalar field-based model, although the framework can apply to a wide range of models. The ingredients of the model are a heavy complex scalar $\Sigma$ which has a large asymmetry and a pair of scalar annihilons $\phi_{B}$ and $\pbh$, of mass $\sim 100 \GeV - 1 \TeV$, oppositely charged under the SM and carrying baryon number, to which $\Sigma$ will decay. We also require a scalar multiplet, $\tH$, to which the annihilons can annihilate. Finally we require operators which allow $\phi_{B}$ and $\pbh$ to decay to a conventional baryon asymmetry. We will not discuss the origin of the initial $\Sigma$ asymmetry, as our concern is its conversion to a thermal WIMP-like baryon density, but we will show that the $\Sigma$ asymmetry can evade thermalization and erasure.

The essential interaction terms are 
\be{e1} {\cal L}_{\Sigma \; decay} =  \frac{1}{M_{*}}\Sigma \phi_{B}^{2} \pbh^2 \;\;\; + \;h. \; c. ~\ee 
and
\be{e2} {\cal L}_{\phi_{B}\pbh \; ann} = \lambda_{B} \phi_{B} \pbh \tH^{\dagger}\tH  \;\;\; + \;h. \; c.  ~.\ee
In principle $\tH$ can be any scalar multiplet. In order to explore the possibility that $\tH$ is the conventional Higgs boson, $H$, we will assume that $\tH$ has the same quantum numbers as $H$. As we will show, 
$\tH$ cannot develop a significant expectation value if the baryon asymmetry is to evade washout. 
In order for the $\phi_{B}$ and $\pbh$ to decay to a conventional baryon asymmetry, we couple them to quarks. We will consider two examples in which $\phi_{B}$ and $\pbh$ carries SM gauge charges: (i) 
$\phi_{B}$ and $\pbh$ carry only hypercharge ($Y(\phi_{B}) = -1$, $Y(\pbh) = 1$); (ii) $\phi_{B}$ and $\pbh$ are colour triplets with hypercharge ($\phi_{B}( {\bf 3}, 2/3)$, $\pbh( {\bf \overline{3}}, -2/3)$). 
For case (i) the operators transferring the baryon number to quarks are
\be{e3a} {\cal L}_{\phi_{B} \; decay} = \frac{1}{M_{D}^3}
 \phi_{B} \psi_{u^c} \psi_{d^c} \psi_{d^c} \psi_{e^c}   \;\;\; + \;h. \; c.  ~\ee
\be{e3b} {\cal L}_{\pbh \; decay} = \frac{1}{M_{D}^3}
 \pbh \psi_{u^c} \psi_{u^c} \overline{\psi}_{Q} \overline{\psi}_{L}   \;\;\; + \;h. \; c.  ~,\ee
where $\psi_{i}$ are two-component spinors and gauge and generation indices are suppressed. 
In this case $B(\phi_{B}) = B(\pbh) = 1$ and $B(\Sigma) = -4$. 
For case (ii) the operators are 
\be{e3c} {\cal L}_{\phi_{B}\; decay} = \frac{1}{M_{D}^3}
 \phi_{B} \psi_{u^c} \psi_{e^c} \psi_{L} \psi_{L}   \;\;\; + \;h. \; c.  ~\ee
\be{e3d} {\cal L}_{\pbh \; decay} = \frac{1}{M_{D}^3}
 \pbh \psi_{u^c} \psi_{d^c} \overline{\psi}_{L} \overline{\psi}_{L}   \;\;\; + \;h. \; c.  ~.\ee
In this case $B(\phi_{B}) = 1/3$, $B(\pbh) = 2/3$ and $B(\Sigma) = -2$. 

     At this point it must be emphasized that there are other couplings which are allowed by the symmetries of the model but which must be suppressed in order to evade B washout or too rapid annihilon decay. In particular, as discussed later, the B-violating mass term $\phi_{B} \pbh$ must be suppressed, which cannot be achieved by any symmetry which permits \eq{e2}. There are also other smaller dimension operators leading to $\phi_{B}$, $\pbh$ decay processes which must be suppressed, for example $\phi_{B} \overline{\psi}_{L}\overline{\psi}_{L}$ in model (i). In addition, we must assume that $\Sigma$ does not couple directly to SM fields. Since there is no symmetry which can explain these suppressions, they must be assumed to occur as a result of the underlying dynamics of a more complete theory. This does not rule out the possibility of a baryomorphosis model which evades the need for ad hoc suppressions. The purpose of the models we are considering here is to illustrate how baryomorphosis models might work and the potential difficulties encountered in implementing baryomorphosis in practice.       

   The sequence of events is as follows. We assume that a large asymmetry in $\Sigma$ scalars exists down to a low temperature ($\lae 100 \GeV$). The $\Sigma$ particles decay at $T_{d}$ to $\phi_{B}$, $\pbh$ pairs via \eq{e1}, resulting in equal $\phi_{B}$ and $\pbh$ asymmetries. We assume that $T_{d}$ is less than the freeze-out temperature of the $\phi_{B}$, $\pbh$ scalars, $T_{\phi_{B}}$, and that the number density of $\phi_{B}$, $\pbh$  produced by $\Sigma$ decay is large enough that the $\phi_{B}$, $\pbh$  density can subsequently annihilate via \eq{e2}. We will assume that $\lambda_{B}$ is of a natural order of magnitude from a dimensional viewpoint, $\lambda_{B} \sim 0.1$. It is essential that $\lambda_{B}$ is not very small compared with 1 in order that the B-violating annihilations are of a similar strength to those of WIMPs. Since the mass of $\phi_{B}$ and $\pbh$ is assumed to be roughly weak scale\footnote{For simplicity we assume 
that the $\phi_{B}$ and $\pbh$ masses are equal, although in general they could have different masses.}, $\sim 100 \GeV- 1 \TeV$, the annihilation cross-section from \eq{e2} is broadly of weak-interaction strength. The resulting $\phi_{B}$ asymmetry therefore corresponds to thermal relic WIMP-like density, although the $\phi_{B}$, $\pbh$ density freezes-out at a value larger than a thermal relic density. At this stage the baryon asymmetry is entirely in the $\phi_{B}$, $\pbh$. To convert this to a conventional baryon asymmetry, the $\phi_{B}$, $\pbh$ density decays at $T_{D} > T_{nuc}$ to quarks via \eq{e3a}-\eq{e3d}, where $T_{nuc} \approx 1 \MeV$ is the temperature of nucleosynthesis.

\section{Thermal WIMP-like Baryon Density via annihilation of annihilons}

\subsection{$\phi_{B}$, $\pbh$ annihilation}    
           
  The non-relativistic annihilation rate times relative velocity for $\phi_{B} \pbh \rightarrow \tH^{\dagger} \tH$ is
 \cite{jmold} 
\be{e5}  \langle \sigma v \rangle  = \frac{ \lambda_{B}^{2}}{16 \pi m^{2}_{\phi_{B}} }     ~.\ee
(Annihilons may be relativistic when first produced but scattering from the SM thermal background will rapidly slow them.) Assuming $\Sigma$ decays to $\phi_{B}$, $\pbh$ at $T_{d}$, annihilations due to \eq{e2} reduce the number density to 
\be{e6} n_{\phi_{B}}(T_{d}) =  \frac{H(T_{d})}{\langle \sigma v \rangle}     ~.\ee 
(A similar density of $\pbh$ will be produced.) 
We can relate this to the thermal relic $\phi_{B}$ density at $T_{d}$ due to $\eq{e2}$ (i.e. the density of thermal relic $\phi_{B}$ scalars in the case when all interactions other then \eq{e2} are neglected), 
\be{e7} n_{\phi_{B}}(T_{d}) = 
\frac{ g^{1/2}(T_{\phi_{B}})T_{\phi_{B}}}{g^{1/2}(T_{d})T_{d}} \times n_{\phi_{B}\;th}(T_{d})   ~,\ee  
where $T_{\phi_{B}}$ is the $\phi_{B}$ freeze-out temperature due to \eq{e2} and we assume that $T_{d} < T_{\phi_{B}}$.
(As usual, $g(T)$ is the effective number of relativistic degrees of freedom.) If the $\phi_{B}$ subsequently decays to conventional baryons, then 
the mass density in baryons (assuming the baryon number of $\phi_{B}$ equals 1) is 
\be{e8} \Omega_{B} h^2 = \frac{m_{n}}{m_{\phi_{B}}} 
\frac{T_{\phi_{B}}}{T_{d}} \times \Omega_{\phi_{B}\; th} h^2   ~,\ee
where $\Omega_{\phi_{B}\; th}$ would be the thermal relic 
$\phi_{B}$ density due to \eq{e2} if they did not decay, which is a typical thermal relic WIMP density if $m_{\phi_{B}} \sim 100 \GeV - 1 \TeV$ and $\lambda_{B} \sim 0.1 - 1$ \cite{jmold}, 
\be{e9} \Omega_{\phi_{B}\; th} = \frac{g(T_{\gamma})}{g(T_{\phi_{B}})} 
\frac{k_{T_{\phi_{B}}} T_{\gamma}^{3}}{x_{\phi_{B}} \langle \sigma v \rangle \rho_{c} M_{Pl}} ~,\ee
where $\rho_{c} = 8.1 \times 10^{-47}h^2 \GeV^{4}$ is the critical density, $x_{\phi_{B}} = T_{\phi_{B}}/m_{\phi_{B}}$ and during radiation-domination $H = k_{T}T^2/M_{Pl}$ where $k_{T} = (4 \pi^{3} g(T)/45)^{1/2}$. ($M_{Pl} = 1.22 \times 10^{19} \GeV$.) Typically $x_{\phi_{B}} \sim 1/20$.   Note that there could be symmetric thermal relic densities of $\phi_{B}$ and $\pbh$, but these are typically small compared to the asymmetric density (since $T_{d} < T_{\phi_{B}}$) and in any case do not contribute to the final baryon number. 
Therefore
\be{e9a} \Omega_{B}h^2 = 0.024 \GeV \times \frac{1}{T_{d}}
\left(\frac{m_{\phi_{B}}}{1 \TeV}\right)^{2}
\left(\frac{0.1}{\lambda_{B}}\right)^{2}
\left(\frac{100}{g(T_{\phi_{B}})}\right)^{1/2} ~.\ee
The observed\footnote{We use the central values of WMAP 7-year mean results: 
$\Omega_{B} = 0.0449 \pm 0.0028$, $\Omega_{DM}h^2 = 0.1334 \pm 0.0056$ and $h = 0.710 \pm 0.025$ \cite{wmap7}.} baryon asymmetry is $\Omega_{B} h^2 = 0.0226$. Therefore to get the correct baryon asymmetry we need 
\be{e10} T_{d} \approx 1.1  \GeV \times
\left( \frac{m_{\phi_{B}}}{1 \TeV} \right)^{2} 
\left( \frac{0.1}{\lambda_{B}} \right)^{2} 
\left(\frac{100}{g(T_{\phi_{B}})}\right)^{1/2}
~.\ee  
This assumes that $T_{d} < T_{\phi_{B}}$, where 
\be{e11} T_{\phi_{B}} = m_{\phi_{B}}x_{\phi_{B}}
 = 50 \GeV  \times \left(\frac{m_{\phi_{B}}}{1 \TeV} \right)
 \left(20 x_{\phi_{B}}\right) ~.\ee    

       Therefore if $\Sigma$ decays to $\phi_{B}$, $\pbh$ at a sufficiently low temperature then the initial $\phi_{B}$ and $\pbh$ asymmetry will annihilate down to a thermal relic WIMP-like density determined by a broadly weak-interaction strength annihilation cross-section. It arises from a process very similar to the freeze-out of annihilations of a thermal equilibrium WIMP density, but here the $\phi_{B}$ density is greater than the thermal equilibrium density when the annihilations freeze-out\footnote{This is similar to the generation of a non-thermal WIMP density from moduli decay, as discussed in \cite{ntwimp}.}. Note that the greater-than-equlibrium  $\phi_{B}$ density can compensate for the small nucleon mass in the final $\Omega_{B}$, allowing $\Omega_{B}$ to become comparable to a typical WIMP dark matter density. 

       The relic baryon asymmetry is typically of the order of magnitude of the observed dark matter density $\Omega_{DM}$. In Fig.1 we show $r_{BDM} \equiv \Omega_{B}/\Omega_{DM}$ as a function of $T_{d}$ and $m_{\phi_{B}}$ for the case where $\lambda_{B} = 0.1$. 
The $\Sigma$ decay temperature must be low enough, $T_{d} < T_{\phi_{B}}$, for the baryon asymmetry to avoid being erased by B-violating $\phi_{B}$ $\pbh$ annihilations. Since the puzzle of the baryon and dark matter densities is why they are within an order of magnitude, we have plotted the region where $0.1 \leq r_{BDM} \leq 10$. From Fig.1 we see that 
a large fraction of the allowed $T_{d}$, $m_{\phi_{B}}$ parameter space has $\Omega_{B}$ and $\Omega_{DM}$ within an order of magnitude. Therefore provided $T_{d}$ is low enough for $\Sigma$ decays to produce a baryon asymmetry (but high enough to allow standard nucleosynthesis), the model naturally explains why the baryon and dark matter mass densities are similar.  

\begin{figure}[htbp]
\begin{center}
\epsfig{file=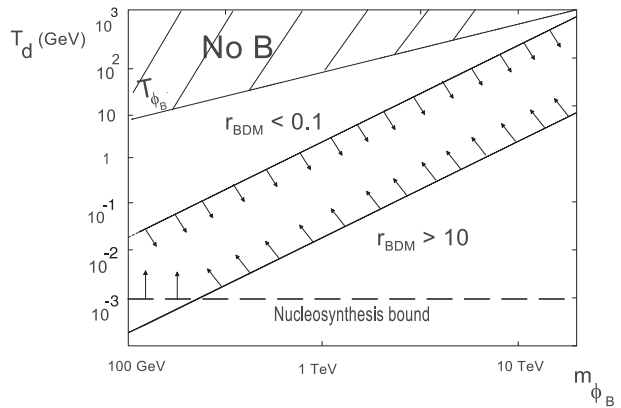, width=0.3\textwidth, angle = -90}
\caption{Region of $T_{d}$, $m_{\phi_{B}}$ parameter space for which $\Omega_{B}$ is within an order of magnitude of $\Omega_{DM}$, for the case $\lambda_{B} = 0.1$. The "No B" region corresponds to 
$T_{d} > T_{\phi_{B}}$, where $\phi_{B}-\pbh$ annihilations are in equilibrium and erase the $B$ asymmetry.}
\label{fig1}
\end{center}
\end{figure}

\subsection{$\phi_{B}$, $\pbh$ decay} 

    The baryon asymmetry in $\phi_{B}$ and $\pbh$ must decay to a conventional baryon asymmetry. This occurs via the operators in \eq{e3a}-\eq{e3d}. The 
$\phi_{B}$, $\pbh$ decay rate to Standard Model (SM) quarks and leptons can be estimated to be 
\be{e12} \Gamma_{\phi_{B}} \approx k_{p} \frac{m_{\phi_{B}}^{7}}{M_{D}^{6}}     ~,\ee
where $k_{p} \approx 1/512\pi^5$ is a kinematical factor for decay to 4-body final states. Decay occurs once $\Gamma_{\phi_{B}} \approx H \equiv k_{T} T^2/M_{Pl}$.
Therefore the mass $M_{D}$ as a function of decay temperaure is 
$$  M_{D} \approx \left(\frac{k_{p}}{k_{T_{D}}} \right)^{1/6}  
\left(\frac{M_{p}}{T_{D}^{2}} \right)^{1/6} m_{\phi_{B}}^{7/6} $$
\be{e14}  = 5 \times 10^{7} \GeV 
\left(\frac{k_{p}}{k_{T}} \right)^{1/6}  
\left(\frac{1 \MeV}{T_{D}} \right)^{1/3}    
\left(\frac{m_{\phi_{B}}}{1 \TeV} \right)^{7/6}   ~.\ee

     The lifetime of the annihilons may provide a signature for such particles at colliders. The $\phi_{B}$ decay temperature satisfies $1 \MeV \lae T_{D} \lae T_{d}$. This range translates into a lifetime $\tau \; (\equiv H^{-1}(T_{D}))$ which is characteristically long,
\be{e14a}   1.5 \; s \; \gae \; \tau \; \gae \; 8 \times 10^{-9} \left( \frac{10 \GeV}{T_{d}}  \right)^2 \; s    ~.\ee 
The lower bound can be short enough to allow annihilons to decay within particle detectors, depending on $T_{d}$.

\section{Constraints}

   In order to have a consistent scenario we must check that (a) the initial $\Sigma$ asymmetry does not come into thermal equilibrium, since in this case the $\Sigma$ asymmetry will be erased by the interaction \eq{e2}, (b) scattering of $\phi_{B}$, $\pbh$ from thermal background SM fermions via \eq{e2} does not erase the baryon asymmetry in $\phi_{B}$, $\pbh$ and (c) that the B-violating interaction \eq{e2} does not induce rapid proton decay.

\subsection{Survival of the $\Sigma$ asymmetry}

      We will consider the $\Sigma$ particles to have zero momentum, as would be the case if they formed a coherently oscillating condensate. This case will give the largest scattering rate of $\Sigma$ by the thermal background $\phi_{B}$ particles, so the bounds obtained will be conservative.

    We first determine the mass in \eq{e1} as a function of the $\Sigma$ decay temperature $T_{d}$. We require that $\Sigma$ decays at $T_{d} < T_{\phi_{B}}$. The $\Sigma \rightarrow \phi_{B} \phi_{B} \pbh \pbh$ decay rate is 
\be{e15} \Gamma_{\Sigma} \approx \frac{k_{p}}{4} \left( \frac{1}{M_{*}}\right)^2 m_{\Sigma}^{3}  ~,\ee
where $k_{p}$ is a 4-body final state factor as before. Decay occurs once $H \approx \Gamma_{\Sigma}$.  
The mass scale $M_{*}$ as a function of $T_{d}$ is therefore 
$$ M_{*} \approx \left( \frac{k_{p}}{k_{T_{d}}} \right)^{1/2} \frac{\left(m_{\Sigma}^{3} M_{p}\right)^{1/2}}{2 T_{d}}$$
\be{e17} 
\;\;\;\;\;\; = 5.5 \times 10^{13} \GeV 
\left(\frac{k_{p}}{k_{T_{d}}}\right)^{1/2} 
\left( \frac{m_{\Sigma}}{1 \TeV} \right)^{3/2}  
\left(\frac{1 \GeV}{T_{d}}\right)  ~.\ee
     
   Assuming that the $\phi_{B}$ particles are in thermal equilibrium and relativistic with $E \approx 3 T > m_{\Sigma}$, $m_{\phi_{B}}$ (the largest scattering rate is given by the highest temperature), the cross-section for scattering of zero-mommentum $\Sigma$ particles via $\phi_{B} \Sigma \rightarrow \phi_{B} \pbh \pbh$ is 
$ \sigma \approx k_{1}/M_{*}^{2}$
where we estimate $k_{1} \approx 1/1024 \pi^3$.
The rate at which $\Sigma$ are scattered is then
\be{e19}  \Gamma_{sc} \approx n_{\phi_{B}}(T) \sigma \approx \frac{k_{1} T^{3}}{\pi^{2} M_{*}^{2}}   ~.\ee 
This must be less than $H$ for all $T$ during radiation-domination i.e. for all $T < T_{R}$, where $T_{R}$ is the reheating temperature. This is true if 
\be{e20} T_{R} \lae \frac{k_{T} \pi^{2} M_{*}^{2}}{k_{1} M_{Pl}}    ~.\ee
Using the relation between $M_{*}$ and $T_{d}$ given by \eq{e17}, this translates into an upper bound on
$T_{R}$, 
\be{e21} T_{R} \lae \left(\frac{k_{p}}{k_{1}}\right) \frac{ \pi^{2} m_{\Sigma}^{3}}{4 T_{d}^{2}}   ~.\ee
Therefore
\be{e21a} T_{R} \lae 5 \times 10^{8} \GeV  
\left( \frac{m_{\Sigma}}{1 \TeV}\right)^{3}
\left( \frac{1 \GeV}{T_{d}} \right)^{2}
~.\ee
Thus the $\Sigma$ asymmetry will not be erased as long as the reheating temperature is sufficiently low\footnote{$H \propto T^4$ while $\Gamma_{sc} \propto T^3$ when $T > T_{R}$, therefore $\Sigma$ are generally out of thermal equilibrium if \eq{e21a} is satisfied.}.

   This demonstrates an interesting conenction between the required low value of $T_{d}$ and the survival of the $\Sigma$ asymmetry: the large value of $M_{*}$ which is necessary to have $T_{d} < T_{\phi_{B}}$ also ensures that $\Sigma$ is out of thermal equilibrium.

\subsection{Evasion of B washout due to $\phi_{B}$, $\pbh$ scattering from thermal background SM fermions.}

    We must ensure that the B-violating interaction \eq{e2} does not lead to washout of the baryon asymmetry in $\phi_{B}$, $\pbh$. In the case where $\tH$ gains an expectation value, the mass squared term due to \eq{e2} results in mixing of $\phi_{B}$ and $\pbh^{\dagger}$. This will cause B-violating scattering $\phi_{B} + \psi \rightarrow \pbh^{\dagger} + \psi$ with the thermal background SM fermions $\psi$ due to t-channel Higgs and gauge boson exchange processes. In addition it will cause oscillations of the initial $\phi_{B} \phi_{B} \pbh \pbh$ state from $\Sigma$ decay to its conjugate\footnote{This aspect of the model has some dynamical similarities to the model of \cite{cohen}, where a dark matter asymmetry initially generated from a lepton asymmetry is subsequently symmetrized by oscillations due to a symmetry breaking mass term.}. We find that the most rapid B-violating process is due to t-channel photon or gluon exchange. Treating the mass squared term due to \eq{e2} as a mass insertion $\Delta m^{2} = \lambda_{B} v^2$, where $<\tH> = v$, we estimate the rate of scattering at $T_{d}$ to be 
\be{x1} \Gamma_{sc} \approx \left(\frac{\Delta m^2}{m_{\phi_{B}}^{2}}\right)^{2} \alpha_{g}^{2}T_{d}    ~,\ee
where $\alpha_{g}= g^2/4\pi$ and $g$ the gauge coupling. 
This is larger than the expansion rate if 
$$ \Delta m \gae \left( \frac{k_{T}}{\alpha_{g}^{2}} \right)^{1/4} \left(\frac{T_{d}}{M_{Pl}}\right)^{1/4} m_{\phi_{B}} $$
\be{x2} = \;  0.1 \GeV \left(\frac{0.1}{\alpha_{g}}\right)^{1/2}
\left(\frac{T_{d}}{1 \GeV}\right)^{1/4}
\left(\frac{m_{\phi_{B}}}{1 \TeV}\right)
~, \ee
where we have used $k_{T} \approx 5$ corresponding to $T_{d} \lae 1 \GeV$. 
For the case where $\tH$ is identified as the Higgs boson, in which case $v = 175 \GeV$, this requires that $\lambda_{B} \lae 3 \times 10^{-7}$. Since this very small value for $\lambda_{B}$ would undermine the weak interaction 
strength of the annihilation process from \eq{e2}, we cannot consider $\tH$ to be the SM Higgs doublet. However, if 
$\tH$ is a second doublet which does not develop an expectation value then the problem of B-violation due to 
mixing will be avoided\footnote{The interaction \eq{e2} will also induce a quadratically divergent 1-loop contribution to $\Delta m^2$, which must be cancelled by counterterms to the level $\Delta m \lae 0.1 \GeV$. This would be a problem in theories where the scalar masses are calculable and are required to be naturally weak scale, such as models with a physical TeV cut-off or weak-scale supersymmetry. But if the cut-off is infinite and the quadratic divergences are unsuppressed, as in the SM, then scalar mass terms cannot be calculated but are instead treated as phenomenological inputs in the renormalization procedure.}. In this case we must ensure that either $\tH$ decays before nucleosynthesis or that the relic density of stable $\tH$ is sufficiently small. One possibility is that $\tH$ could decay to SM fermions via Yukawa-like couplings. Alternatively, if $\tH$ is stable (for example if there is a $Z_{2}$-symmetry $\tH \leftrightarrow -\tH$), then it would be a candidate for inert doublet dark matter \cite{iddm}. This would have the advantage of introducing a WIMP candidate into the SM while simultaneously accounting for a WIMP-like baryon asymmetry\footnote{We could also consider gauge singlet dark matter scalars as a final state in \eq{e2}.}. We will return to this possibility in future work.

\subsection{Proton stability}

   The final requirement is that the B-violating interaction \eq{e2} does not lead to rapid proton decay. The mass scales of the particles and of the interactions \eq{e3a}-\eq{e3d} are relatively low, so we should check that rapid proton decay does not occur due to the combination of \eq{e2} and \eq{e3a}-\eq{e3d}. The strongest effective B-violating quark interaction which can lead to proton decay
is obtained by introducing a mass splitting $\Delta m^2$ for $\phi_{B}$, $\pbh$ (where $\Delta m \lae 0.1 \GeV$ in order to evade B washout) and integrating out the heavy $\phi_{B}$ and $\pbh$, as in Fig.2. For the model where $\phi_{B}$, $\pbh$ are colour singlets with hypercharge this gives 
\be{e22} {\cal L}_{eff} = \frac{\Delta m^2}{M_{D}^{6} m_{\phi_{B}}^4}
 \left( \psi_{u^{c}} \psi_{d^{c}}  \psi_{d^{c}} \psi_{e^{c}} \right)
 \left(\psi_{u^c} \psi_{u^c} \overline{\psi}_{Q} \overline{\psi}_{L} \right)
~.\ee    
From this we estimate the proton decay rate dimensionally to be
\be{e23} \Gamma_{proton} \approx k_{\Gamma} \left|\frac{\Delta m^2}{M_{D}^{6} m_{\phi_{B}}^4}\right|^{2} M^{17}   ~,\ee  
where $k_{\Gamma} \ll 1$ is a factor representing phase space and hadron physics. (This assumes that proton decay via \eq{e22} is not kinematically suppressed, so it is likely to be an overestimate.)     
The mass scale $M$ is expected to be set by the nucleon mass $m_{n}$. 
Replacing $M_{D}$ by its value in \eq{e14} and setting $M = m_{n}$ gives
\be{e24} \tau = \frac{1}{k_{\Gamma}} \left( \frac{k_{p}}{k_{T_{D}}} \right)^{2} 
\frac{ m_{\phi_{B}}^{22} M_{Pl}^{2}}{ T_{D}^{4} \Delta m^{4} m_{n}^{17}} ~.\ee
Therefore 
\be{e25} \tau \approx 10^{88} yrs \times  
\frac{1}{k_{\Gamma}}
\left( \frac{0.1 \GeV}{\Delta m} \right)^{4}
\left( \frac{k_{p}}{k_{T_{D}}} \right)^{2} 
\left( \frac{m_{\phi_{B}}}{1 \TeV} \right)^{22}
\left( \frac{1 \MeV}{T_{D}} \right)^{4} 
~.\ee
Therefore, in spite of the extreme sensitivity to $m_{\phi_{B}}$, the proton is sufficiently stable in this model.

\begin{figure}[htbp]
\begin{center}
\epsfig{file=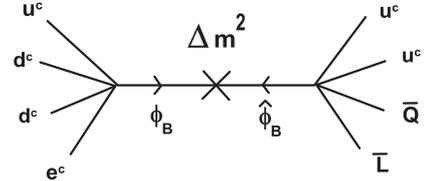, width=0.3\textwidth, angle = 0}
\caption{Operator leading to proton instability via $\phi_{B}$, $\pbh$ exchange.}
\label{fig2}
\end{center}
\end{figure}

\section{Experimental Signatures}

   The characteristic feature of these models are the annihilons which carry baryon number and annihilate to a WIMP-like thermal relic baryon density.  Annihilons in general would be long-lived massive particles, with typically $\tau \gae 10^{-8} s$. The prospect for producing and identifying these at colliders depends on their SM gauge charges. The best prospect would be the case where the annihilons are coloured scalars. In this case they could be pair produced at the LHC via gluon fusion\footnote{Their production would be similar to scalar leptoquarks and gluons, which can be pair produced at the LHC for masses up to about 1 TeV \cite{scalarg}.}. Coloured annihilons would combine with quarks to form heavy hadron-like particles with unusual charge and baryon number assignments. Their decay to baryon number could be observed if the annihilons could decay or be stopped within a detector. For example, for the coloured scalar annihilon example discussed in this paper, the $\pbh$ can be produced as heavy $\pbh$-hadrons e.g. $d\;\pbh$, which has $B = 1$ and $Q = -1$. Such long-lived exotic hadrons (with various possible charge and baryon number assignments) are a generic prediction of baryomorphosis models in the case where the annihilons are coloured. 

     It may also be possible to produce and detect annihilons at colliders in the case where they carry only electroweak quantum numbers. We will return to annihilon production and detection at colliders in more detail in a future discussion.

\section{Conclusions}

    The origin of the observed baryon asymmetry is a fundamental question in cosmology. In particular, the similarity of the baryon and dark matter mass densities may be a hint as to the origin of both densities. However, existing models which relate these densities neglect the other well-known coincidence, the similarity of the observed dark matter density to a typical thermal relic WIMP density, the so-called WIMP miracle. An alternative to directly connecting the dark matter and baryon densities is to indirectly relate the observed baryon density to the thermal relic WIMP density i.e. to have similar but seperate mechanisms for the origin of the observed densities rather than directly connecting them to a common origin. This also has the advantage that it does not restrict the mass of the dark matter particle to be light. In this paper we have shown how this might be done, by modifying an initial large baryon asymmetry to a final asymmetry determined by a WIMP-like baryon number-violating interaction, a framework we have termed "baryomorphosis". This generically requires new particles in the 100 GeV - 1 TeV range, "annihilons",  which carry baryon number and which can annihilate to reduce the initially large baryon asymmetry to a thermal relic WIMP-like density.

    This framework requires a number of conditions to be satisifed, in particular the low temperature at which the initial asymmetry is transferred to annihilons and the late decay of annihilons to 
quarks. The low value of $T_{d}$ is phemonemologcally essential, since if $T_{d}$ is greater than the freeze-out temperature of the B-violating annihilation process then the baryon asymmetry will be completely erased. We have noted an encouraging consistency between the low temperature of annihilon decay and the requirement that the initial baryon asymmetry is not erased by B-violating annihilation processes.  Although the final baryon to dark matter density ratio is sensitive to $T_{d}$, the mass of the annihilon and the strength of the B-violating interaction, we find that 
a large fraction of the $T_{d}$, $m_{\phi_{B}}$ parameter space allowed by non-erasure of the baryon asymmetry and by nucleosynthesis is consistent with $\Omega_{B}$ and $\Omega_{DM}$ being within an order of magnitude. Therefore the model can naturally account for the observed value of $\Omega_{B}/\Omega_{DM}$ provided $T_{d}$ is low enough and $\phi_{B}$, $\pbh$ can decay to quarks before nucleosynthesis.

   The model we have presented may also be considered as an illustration of the conditions necessary to explain a thermal WIMP-like baryon density and so the plausibility of explaining this feature of the observed Universe mechanistically. If the conditions appear implausible, then the possible conclusions are (i) the "WIMP miracle" explanation of dark matter is correct but no mechanistic connection with the baryon density is possible, in which case anthropic selection must determine the baryon asymmetry, or (ii) the baryon asymmetry and dark matter density are mechanistically related to each other, but the "WIMP miracle" is not the explanation of the dark matter density. (We are discounting the possibility that the similarity of the baryon and dark matter densities is just a coincidence.) From our model it is clear that some highly non-trivial features and conditions are required: the injection of a large primordial baryon asymmetry at low temperature $\lae 1 \GeV$, the evasion of B washout due to the B-violating annihilation process, and the long lifetime of the annihilons (long enough for them to survive until the time of asymmetry injection but short enough to allow them to decay before nucleosynthesis). In particular, a number of mass terms and couplings which are allowed by the symmetries of the model must be suppressed.  Should such conditions be difficult to satisfy in a natural particle physics model, it would be an indication that either the similarity of the thermal relic WIMP density to the observed dark matter density is a coincidence with no physical significance or that anthropic selection of the baryon asymmetry must play a fundmental role in determining the state of the Universe at present. 
 
     Experimental tests of such models are important in order to eliminate the need for aesthetic judgements regarding their naturalness. We have noted that the annihilons will have properties which may allow them to be identified if they can be produced at colliders, in particular a long lifetime $\gae 10^{-8}$ s. Most promising would be the case of coloured annihilons, which would form heavy hadron-like states with unusual combinations of charge and baryon number, which might either be stopped and their decay observed, or possibly decay within a collider detector.

    We emphasize that the simple scalar field models we have presented here are intended to introduce a general framework and to understand the conditions necessary to implement it. We expect that this framework, baryomorphosis, can be realized in a wide range of models, some of which may be able to address the naturalness issues encountered in the simple models presented here.

\section*{Acknowledgements} 
This work was supported by the European Union through the Marie Curie Research and Training Network "UniverseNet" (MRTN-CT-2006-035863).



\begin{thebibliography}{99}


\bibitem{bdm0}  S.M.Barr, R.S.Chivukula and E.Fahri, \pll{B241}{1990}{387};
 S.M.Barr, \prdd{D44}{1991}{3062}; D.B.Kaplan, \prll{68}{1992}{741}.

\bibitem{bdm1} N.Cosme, L.Lopez Honorez and M.H.G.Tytgat, \prdd{D72}{2005}{043505}; 
R.Kitano and I.Low, \prdd{D71}{2005}{023510}; R.Kitano and I.Low,
  arXiv:hep-ph/0503112.
M.~Y.~Khlopov,
  Pisma Zh.\ Eksp.\ Teor.\ Fiz.\  {\bf 83}, 3 (2006)
  [JETP Lett.\  {\bf 83}, 1 (2006)]
  [arXiv:astro-ph/0511796].
K.~Agashe and G.~Servant,
  JCAP {\bf 0502}, 002 (2005)
  [arXiv:hep-ph/0411254];
D.~Hooper, J.~March-Russell and S.~M.~West,
  Phys.\ Lett.\  B {\bf 605} (2005) 228
  [arXiv:hep-ph/0410114];
 G.~R.~Farrar and G.~Zaharijas,
  Phys.\ Rev.\ Lett.\  {\bf 96}, 041302 (2006)
  [arXiv:hep-ph/0510079];
 S.~Abel and V.~Page,
  JHEP {\bf 0605}, 024 (2006)
  [arXiv:hep-ph/0601149].
H.~An, S.~L.~Chen, R.~N.~Mohapatra and Y.~Zhang,
  JHEP {\bf 1003}, 124 (2010)
  [arXiv:0911.4463 [hep-ph]].


 \bibitem{susybdm0} S.D.Thomas, \pll{B356}{1995}{256}. 


\bibitem{susybdm1} K.Enqvist and J.McDonald, \npp{B538}{1999}{321};
 K.~Enqvist and J.~McDonald,
  Phys.\ Lett.\  B {\bf 440}, 59 (1998)
  [arXiv:hep-ph/9807269].
  L.~Roszkowski and O.~Seto,
  Phys.\ Rev.\ Lett.\  {\bf 98}, 161304 (2007)
  [arXiv:hep-ph/0608013];
  O.~Seto and M.~Yamaguchi,
  Phys.\ Rev.\  D {\bf 75}, 123506 (2007)
  [arXiv:0704.0510 [hep-ph]].


\bibitem{x1}   Y.~Cai, M.~A.~Luty and D.~E.~Kaplan,
  arXiv:0909.5499 [hep-ph].


\bibitem{xo} 
  M.~R.~Buckley and L.~Randall,
  arXiv:1009.0270 [hep-ph].

\bibitem{x2}  P.~H.~Gu, M.~Lindner, U.~Sarkar and X.~Zhang,
  arXiv:1009.2690 [hep-ph].


\bibitem{bdmr}
  T.~Cohen, D.~J.~Phalen, A.~Pierce and K.~M.~Zurek,
  Phys.\ Rev.\  D {\bf 82}, 056001 (2010)
  [arXiv:1005.1655 [hep-ph]];
  H.~Davoudiasl, D.~E.~Morrissey, K.~Sigurdson and S.~Tulin,
  arXiv:1008.2399 [hep-ph];
  J.~Shelton and K.~M.~Zurek,
  arXiv:1008.1997 [hep-ph];
M.~Blennow, B.~Dasgupta, E.~Fernandez-Martinez and N.~Rius,
  arXiv:1009.3159 [hep-ph].


\bibitem{jm1}
 J.~McDonald,
  JCAP {\bf 0701}, 001 (2007)
  [arXiv:hep-ph/0609126].


\bibitem{jmold}
J.~McDonald,
  Phys.\ Rev.\  D {\bf 50}, 3637 (1994)
  [arXiv:hep-ph/0702143].


\bibitem{wmap7}  E.~Komatsu {\it et al.},
  arXiv:1001.4538 [astro-ph.CO].

\bibitem{ntwimp}
  B.~S.~Acharya, G.~Kane, S.~Watson and P.~Kumar,
  Phys.\ Rev.\  D {\bf 80}, 083529 (2009)
  [arXiv:0908.2430 [astro-ph.CO]].


\bibitem{cohen}
  T.~Cohen and K.~M.~Zurek,
  Phys.\ Rev.\ Lett.\  {\bf 104}, 101301 (2010)
  [arXiv:0909.2035 [hep-ph]].



\bibitem{iddm}
  L.~Lopez Honorez, E.~Nezri, J.~F.~Oliver and M.~H.~G.~Tytgat,
  JCAP {\bf 0702}, 028 (2007)
  [arXiv:hep-ph/0612275].



\bibitem{scalarg}
  M.~V.~Martynov and A.~D.~Smirnov,
  Mod.\ Phys.\ Lett.\  A {\bf 23}, 2907 (2008)
  [arXiv:0807.4486 [hep-ph]].


\end{thebibliography}
\end{document}